\documentclass{nature-preprint}
\usepackage{tcolorbox,color}
\usepackage{colortbl}
\usepackage{ulem}
\usepackage{amsmath}
\usepackage{amssymb}
\usepackage{graphicx}
\usepackage{upgreek}

\usepackage{caption,wrapfig}
\usepackage{hyperref}
\usepackage{lipsum}

\usepackage[super,comma,compress]{natbib}

\title{Photo memtransistor based on CMOS flash memory technology on Graphene with neuromorphic applications}
\author{Christian~Frydendahl$^{1,\dagger,*}$, S.R.K.~Chaitanya~Indukuri$^{1,\dagger}$, Meir~Grajower$^{1,2}$, Noa Mazurski$^{1}$, Joseph~Shappir$^{1}$, and Uriel~Levy$^{1,*}$}

\begin{document}

\maketitle 

\begin{affiliations}
\small
\item Department of Applied Physics, The Faculty of Science, The Center for Nanoscience and Nanotechnology, The Hebrew University of Jerusalem, Jerusalem 91904, Israel.
\item Thomas J. Watson Laboratory of Applied Physics, California Institute of Technology, Pasadena, CA 91125, USA.
\item[]
\item[$\dagger$] These authors contributed equally to this work
\item[*] christia.frydendahl1@mail.huji.ac.il and ulevy@mail.huji.ac.il
\end{affiliations}


\begin{abstract}
Graphene holds a great promise for a number of diverse future applications, in particular related to its easily tunable doping and Fermi level by electrostatic gating. However, as of today, most implementations rely on electrical doping via the application of continuous large voltages to maintain the desired doping. We show here how graphene can be implemented with conventional semiconductor flash memory technology in order to make programmable doping possible, simply by the application of short gate pulses. We also demonstrate how this approach can be used for a memory device, and also show potential neuromorphic capabilities of the device. Finally, we show that the overall performance can be significantly enhanced by illuminating the device with UV radiation. Our approach may pave the way for integrating graphene in CMOS technology memory applications, and our device design could also be suitable for large scale neuromorphic computing structures.
\end{abstract}


\section*{Introduction}
Graphene, a one atom thick layer of carbon, is promising a future revolution in electronics due to its extraordinary  properties. many graphene based devices have already been realized, such as field effect transistors\cite{Schwierz:2010}, photodetectors\cite{Bonaccorso:2010,Koppens:2014,Goykhman:2016}, photovoltaic solar cells\cite{Yin:2014}, memory devices\cite{Hong:2011,Choi:2013,Vu:2016} etc, while technologies such as touch screens\cite{Bae:2010} and batteries are already seeing commercialization\cite{Kong:2019}.

While significant attention in recent years has been paid towards making 'next-generation' devices consisting of purely 2D materials in the race for superior performance\cite{Hong:2011,Choi:2013,Koppens:2014,Vu:2016}, it is still critical to find implementations of graphene/2D materials based on traditional Complementary Metal-Oxide-Semiconductor (CMOS) technology\cite{Indukuri:2020}. CMOS is by far the most mature and widespread platform currently available, and it stands to reason that if graphene is to fully make the leap from the laboratory to consumer grade devices, it may first have to be through CMOS related fabrication processes and devices.

Motivated by this reasoning, we present here a method for integrating graphene with conventional flash memory technology to enable programmable doping of graphene by electrical gate pulses in a memory transitor (memtransitor) configuration. Furthermore, we show how the efficiency of the charge trapping process responsible for the doping can be increased significantly by the application of ultraviolet (UV) light, and how UV can be used to erase the memory state as well. Graphene based flash memory devices were previously proposed based on photon doping by purely 2D material stacks\cite{lei2014optoelectronic,bertolazzi2013nonvolatile}, trapping and detrapping of charge carries between graphene and oxide interfaces\cite{meng2013ultraviolet}, and more recently by the trapping of charges in low-high band gap dielectric interfaces\cite{choi2017photoelectric}. However, due to a lack of control over the surface trap states, the low degree of reproducibility of these types of devices pose significant challenges for their integration as a large number of memory cells together in a combined device\cite{cao2014can}.

Finally, multi-terminal memtransitors are becoming an important building block in future neuromorphic computing architectures\cite{Sangwan:2018,Sangwan:2020,Markovic:2020}. Neuromorphic computing has many promising future prospects, chief among them is performing low power consumption and high speed computational tasks, in particular image and speech recognition. Training a neural algorithm for speech recognition on standard computing hardware can consume in excess of 1000\,kWh of energy, which is enough to power a human brain for up to six years for all of its general tasks\cite{Markovic:2020}. There is thus a great interest in developing computing architectures that try to emulate the actual functioning of a biological brain. Our approach has the potential to be adopted for such applications. 

Hereby, we report a controlled electrostatic and optical doping of graphene by integrating it with well-established CMOS compatible MONOS (Metal-Oxide-Nitride-Oxide-Silicon) non-volatile flash memory technology, and we also investigate briefly how our device performs as an artificial neuron and test its synaptic plasticity. Our device is expected to be useful as a CMOS compatible building block in a potential large scale neural network, or in general for devices that exploit controlled long-term doping of graphene for a multitude of optical or electrical applications, i.e. analog or digital memory devices.


\section*{Results}

\subsection{Device design and fabrication:}

\begin{figure*}[t]
    \centering
	\includegraphics[width=0.99\linewidth]{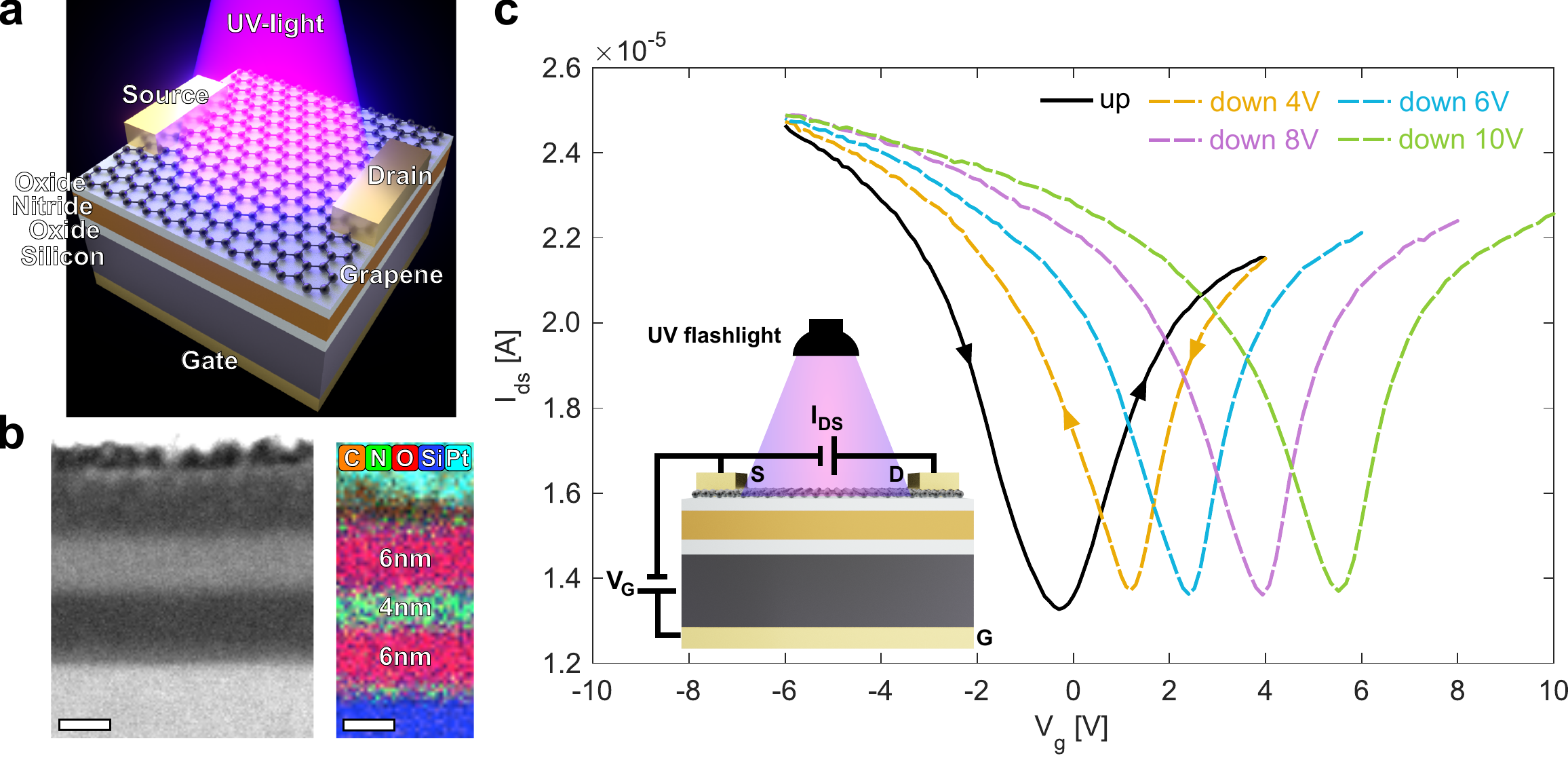}
    \caption{\textbf{a)} Schematic of the investigated graphene memory device. \textbf{b)} STEM dark field image and x-ray elemental composition image showing the ONO stack and its elemental composition. The graphene layer is faintly visible on the top oxide layer. The platinum layer on top of the stack is needed during the FIB process to cut out the sample for STEM imaging, and is not related to the real device geometry. Both scale bars are 5\,nm. \textbf{c)}  Sequential $I_\text{ds}$ vs $V_\text{g}$ sweeps. Sweeping from -6\,V to $V_\text{max}$ (up) and then from $V_\text{max}$ to -6\,V (down), with $V_\text{max} = 4, 6, 8, 10$\,V. A clear hysteresis in $I_\text{ds}$ is seen between up and down, and the hysteresis grows for increasing values of $V_\text{max}$. Insert shows the electrical configuration used during measurements.}
    \label{fig:1}
\end{figure*}

Our device design is as follows: A graphene sheet made using a standard mechanical exfoliation process is dry transferred on top of a layered heterostructure of silicon oxide, silicon nitride, and silicon oxide (ONO) by a standard viscoelastic polymer transfer method\cite{pizzocchero:2016}. The bottom substrate is p-type silicon. The top oxide layer of the ONO has a thickness of 6\,nm, the middle nitride layer is 4\,nm, and the bottom oxide layer is 6\,nm. Three electrical terminals are attached to allow for electrostatic gating perpendicular to the graphene layer, as well as to measure a current in parallel through the graphene sheet. Our design thus follows a standard backgated graphene transistor\cite{Schwierz:2010}, with the exception of the additional nitride layer in between the two oxide layers. The nitride layer is used for charge trapping, acting as a floating gate. A schematic of the device can be seen in Fig.~\ref{fig:1}.a.

To allow for numerous  memory cells to be created, we used the local oxidation of silicon (LOCOS) method to planarize the surface beneath the graphene. This process results in a thick ($\sim$270 nm) and planar oxide layer formed between any two neighboring devices. This reduces cross talk between the close memory cells and allows for integration of a large number of cells. The full fabrication flow is described in the methods section below.

While the top and bottom oxides are of similar thicknesses, the bottom oxide is created by dry thermal oxidation of silicon, while the top oxide is made by a wet oxidation of the nitride. The result is that the bottom oxide is a stronger insulator, and thus the majority of charge carrier exchange happens between the graphene layer and the nitride through the top oxide layer. 

\begin{figure*}[h]
    \centering
	\includegraphics[width=0.8\linewidth]{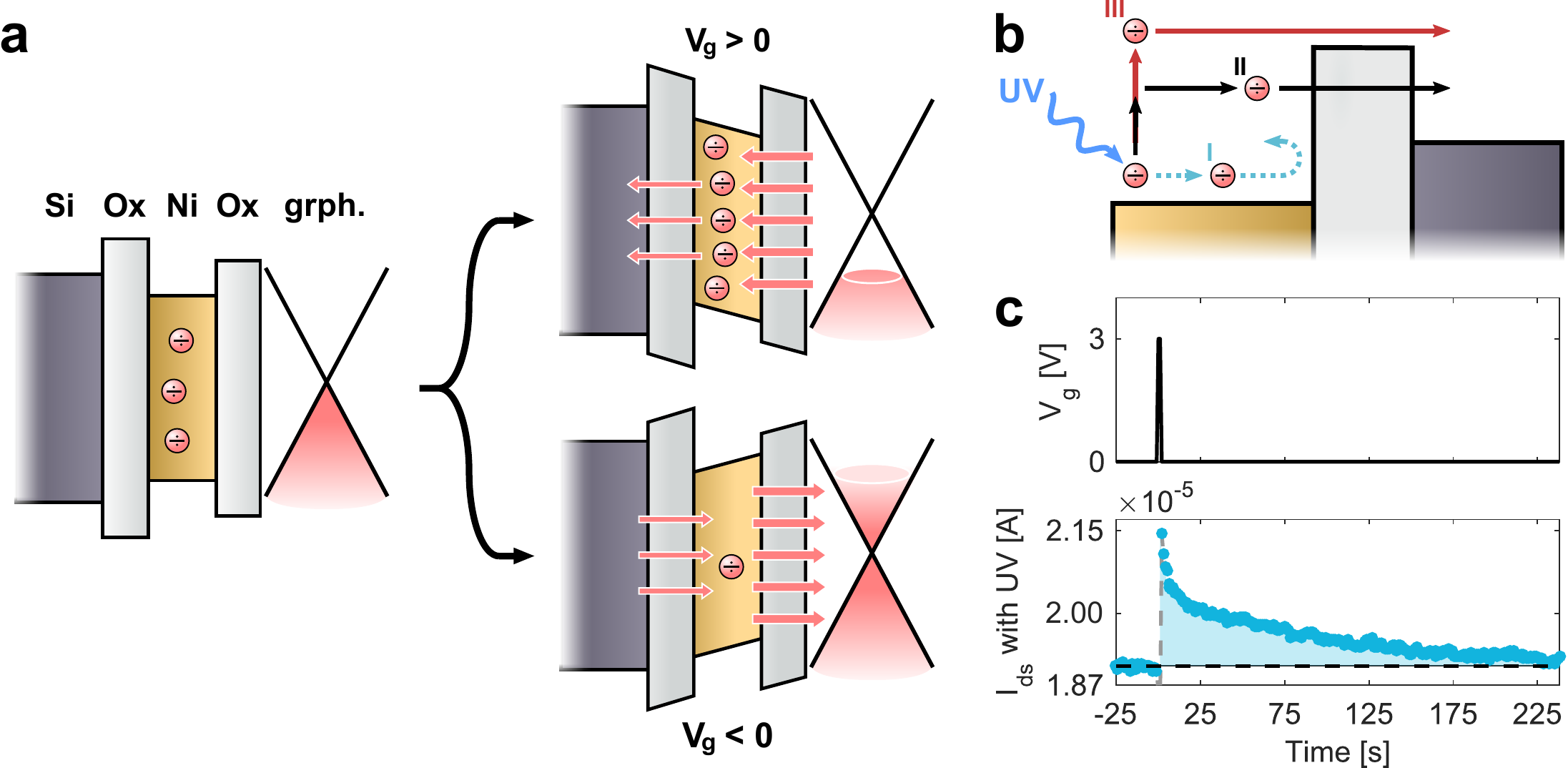}
    \caption{\textbf{a)} Schematic of the tunneling and charge trapping process. For positive gate voltages, electrons are trapped from the graphene into the nitride. This causes p-doping of the graphene. For negative gate potentials, electrons are detrapped from the nitride into the graphene, causing n-doping of the graphene. \textbf{b)} Schematic of how UV light can detrap electrons in the nitride. A carrier with too little energy to escape will be reflected by the interface (I), but if it absorbs a UV photon its energy is increased, allowing the carrier to escape the nitride either by tunneling (II) or directly emission over the barrier (III). \textbf{c)} Optical detrapping of charges from the nitride. A positive gate potential is applied causing an initial rise in the graphene channel's conductivity, but sustained UV illumination on the sample slowly permits the trapped charges in the nitride to escape again.}
    \label{fig:2}
\end{figure*}

Following their fabrication, the devices were measured electrically using a micro-mechanical probe system, see methods below. By applying a positive or negative voltage across the source and gate terminals, $V_\text{g}$, we can modify the conductivity of the graphene sheet taking advantage of the field effect\cite{Schwierz:2010,Xiao:2016}. As a result, the current measured across the source and drain terminals, $I_\text{ds}$ will be changed as well. Fig.~\ref{fig:1}.c shows the measured $I_\text{ds}$ for sweeping $V_\text{g}$ up from -6\,V to $V_\text{max}= 4, 6, 8, 10$\,V, and then down from the respective $V_\text{max}$ to -6\,V. $I_\text{ds}$ is measured by applying a constant voltage, $V_\text{ds}$, of 0.1\,V across the source and drain terminals. The results show a clear hysteresis in the device operation, with the charge neutrality point of the graphene being significantly shifted for the downward sweep compared to the upward sweep (the minimum of the curve), and we can also observe that the hysteresis grows for increasing values of $V_\text{max}$. We see generally that after the first up/down sweep, the device stabilizes into reproducible results. This is most likely because of releasing of trapped charge states/defects in the ONO layer's 'initial state', and is commonly seen for such devices\cite{Emboras:2013}.

The hysteresis seen in the $I_\text{ds}$ vs $V_\text{g}$ characteristics serves as evidence of charge trapping into vacant states in the nitride layer in the ONO stack\cite{Pavan:1997}. We confirm this by fabricating a control device with 25\,nm silicon oxide separating the graphene and the silicon, in which we see a much smaller hysteresis. This small hysteresis is attributed to the vacancy between the graphene-oxide interface. Some charge trapping will always occur in oxide layers, but the presence of the silicon nitride in our main device enhances this charge trapping effect significantly\cite{Pavan:1997}.

By applying a vertical bias across the graphene and ONO layer, charges can tunnel into vacant states in the nitride layer\cite{Pavan:1997}. Here these charges will accumulate and exert a continuous electric field on the graphene layer, resulting in a change in the graphene's Fermi level/charge carrier density, thus changing its conductivity. A similar technique was recently exploited for post-fabrication tuning of silicon photonic ring resonators\cite{grajower:2018}. When a positive bias is applied on the gate electrode, electrons will tunnel from the graphene into the nitride layer. The result is p-doping of the graphene, seen in the top half of the schematic in Fig.~\ref{fig:2}.a. Likewise, when a negative bias is applied, electrons from the nitride will detrap to the graphene, resulting in n-doping of the graphene as shown in the bottom of Fig.~\ref{fig:2}.a. 

Furthermore, it is also possible to detrap trapped electrons by using ultraviolet (UV) light. A schematic of the concept can be seen in Fig.~\ref{fig:2}.b. By sending a short positive gate pulse and keeping the device under constant UV illumination (Fig.~\ref{fig:2}.c, showing the measured current), we can see how the conductivity initially goes up instantaneously after the application of the pulse, but then drops off back gradually to the initial level before the trapping event.

\begin{figure*}[h]
    \centering
	\includegraphics[width=0.8\linewidth]{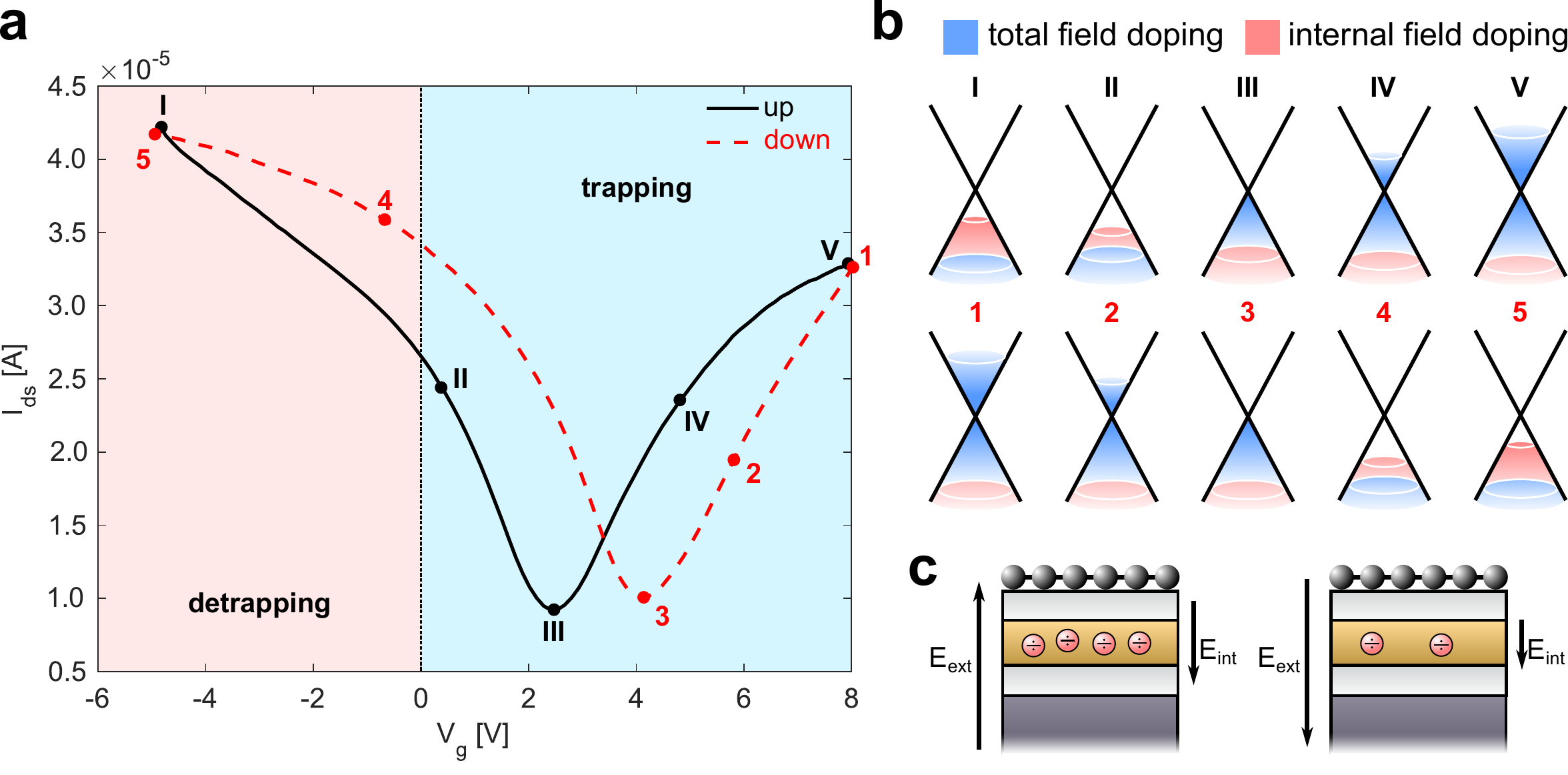}
    \caption{\textbf{a)} Hysteresis of $I_\text{ds}$ vs $V_\text{g}$. \textbf{b)} Graphene band structure for the discrete points in \textbf{a}. Roman numerals refer to the upward portion of the sweep, and the arabic numerals refer to points on the downward portion of the sweep. We distinguish here between the total field effect doping, and the doping effect coming purely from the charges trapped in the nitride. See main text for full explanation. \textbf{c)} Schematic of how the charges trapped in the the nitride contributes to an internal field that dopes the graphene in addition to the externally applied gate bias.}
    \label{fig:3}
\end{figure*}

We can now fully explain the behavior of the hysteresis of the $I_\text{ds}$ vs $V_\text{g}$ measurement. If we refer to Fig.~\ref{fig:3}.b, we can see that initially in point I before starting the $V_\text{g}$ sweep, the graphene is intrinsically only slightly p-doped, but the strong negative bias results in a total doping which makes the graphene strongly p-doped. As the bias now shifts to positive values in point II, we see that the nitride now starts to accumulate charge, making the internal field doping tend more towards p-type, while the positive gate bias is now beginning to move the total field doping of the graphene towards n-doping. The internal field created by the charges in the nitride and how it relates to the externally applied field can be seen in Fig.~\ref{fig:3}.c. In point III we have reached charge neutrality, and we see that the internal field doping has gone more towards p-type again because of the positive applied gate, and as a result a positive bias is needed to counteract it to see charge neutrality. As we increase the positive gate through points IV and V, we see that the internal doping progressively gets pushed more towards p-type due to continued charge accumulation, while the larger applied gates overpowers it and make the graphene solidly n-type. As we now start to decrease the gate again through points 1, 2 and 3, we see that the number of charges stored in the nitride does not change much, as the charges already present will screen out additional charges being added, as the lower gate potentials do not add sufficient energy to add more charges to the nitride. Because of the additional charges stored in the nitride, we also see the charge neutrality point at an even large positive bias than for point III (i.e. more n-doping must be applied to counteract the internal p-doping). This is also why we see a greater shift to p-type for larger values of $V_\text{max}$ in Fig.~\ref{fig:1} (as a larger $V_\text{max}$ results in more charge accumulated in the nitride). When we then cross over to negative bias in point 4, we see that now the nitride begins to discharge rapidly, and when arriving at point 5 we have returned more or less to the initial state of the device.

\subsection{Graphene flash memory:} 
\begin{figure}[h]
    \centering
	\includegraphics[width=0.5\columnwidth]{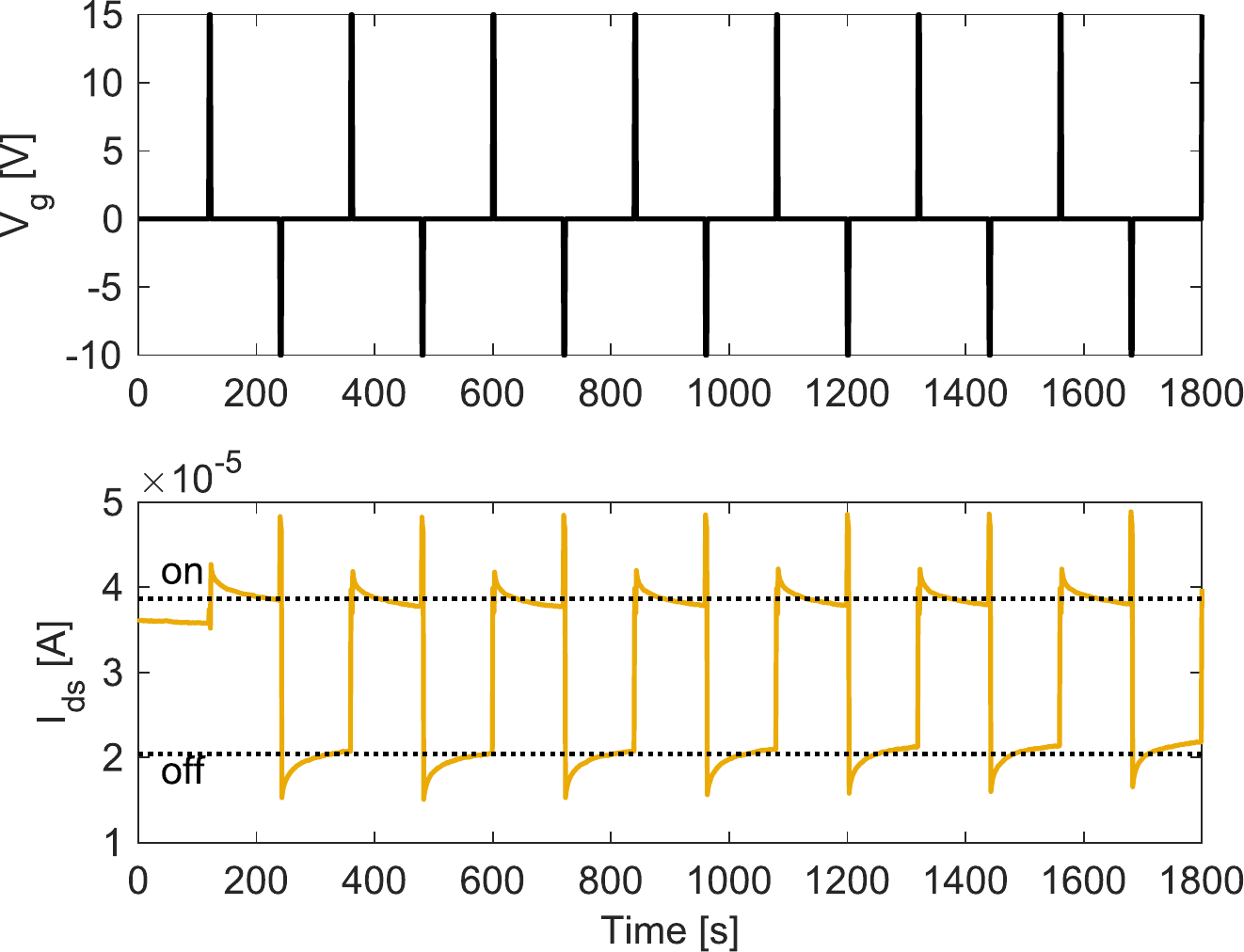}
    \caption{$I_\text{ds}$ as a function of pulsed $V_\text{g}$, here using 2\,s long pulses. The 'off' state is defined as the low conductive state, and the 'on' state as the high conductive state. The large spikes in the current relate to when the gate pulses are engaged, and are caused by the much larger total field exerted on the sample than just the internal field from the nitride after the pulse.}
    \label{fig:4}
\end{figure}

To investigate how this charge trapping can further be exploited for memory applications (as well as memory based graphene doping), we analyze our device by applying pulsed gates for 2\,s alternating between 15\,V and -10\,V, see Fig.~\ref{fig:4}. 

We can now define the higher conductive state after the negative gate pulse as the 'on' state, while the lower conductive state after the positive gate pulse is defined as the 'off' state. We can define the On/Off ratio of the device as a figure of merit:
\begin{equation}
    \text{On}/\text{Off} = \frac{I_\text{on}}{I_\text{off}},    
\end{equation}
with $I_\text{on}$ and $I_\text{off}$ being the high and low current levels read halfway after and before the next gate pulse. From Fig.~\ref{fig:4} we see that an On/Off ratio of $\sim1.8$ is achieved. 

\subsection{Neuromorphic applications:} Finally, due to the memory effect associated with our device, we investigate if the fabricated memtransistor could become relevant for neuromorphic computing applications. By 'neuromorphic' it is generally understood that the transistor should behave as an organic neuron, i.e. it should exhibit 'synaptic/neural plasticity', or in other words, its conductivity should change with its usage\cite{Sangwan:2018,Sangwan:2020,Xu:2016,Van:2017}. Synaptic plasticity is generally categorized in the categories of short- and long term plasticity, with the difference being between the timescales of the induced change.

Short term plasticity refers to changes induced in the device's conductivity for periods of fractions of a second to minutes, and in a real brain is related to the encoding of temporal information in auditory/visual signal processing. This is relevant in associative learning, information processing, pattern recognition, and sound-source localization\cite{Xu:2016}. Long term plasticity refers to more or less permanent changes in the synaptic strength/device conductivity. In a real brain, memories are generally understood to be stored as long term modifications of synaptic strength in the network of neurons, and as such long term plasticity is generally believed to be important for long term memory and learning\cite{Xu:2016}.  

\begin{figure}[h]
    \centering
	\includegraphics[width=\columnwidth]{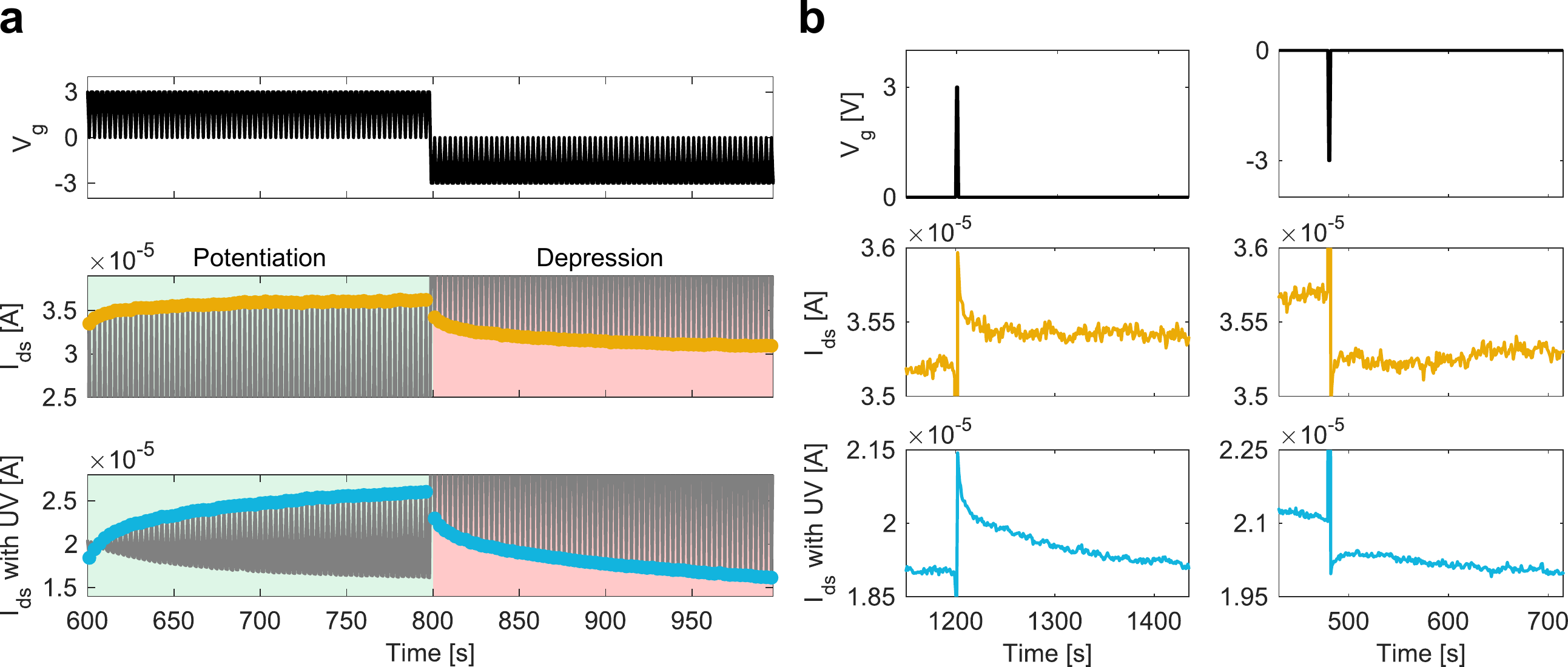}
	\caption{\textbf{a)} Short term potentiation and depression of the graphene memtransistor. The device is spiked in groups of 100 positive, followed by 100 negative spikes. Here shown just a single period of the full sequence. \textbf{b)} Long term potentiation and depression of the same memtransistor by applying a single spike (positive spike on the left panels, negative spike on the right panels), and monitor the resulting $I_\text{ds}$ without and with UV light over time. For both the short- and long term experiments, 2s gate spikes of 3/-3\,V were used. As in previous measurements $V_\text{ds} = 0.1$\,V. Both experiments were done with and without UV illumination.}
    \label{fig:5}
\end{figure}

We characterize the short- and long term plasticity of the device as follows. The short term plasticity is studied by sending a series of gate potential pulses, or in the terminology of neuroscience 'presynaptic spikes', with 1\,s between them. The resulting $I_\text{ds}$, or 'postsynaptic current', can be seen in Fig.~\ref{fig:5}.a. We send first 100 positive spikes to increase the device conductivity (potentiation), followed by 100 negative spikes to lower the device conductivity (depression). To characterize the long term plasticity, we send just a single potential spike, and monitor how the postsynaptic current evolves with time afterwards, see Fig.~\ref{fig:5}.b.

We repeated both the short- and long term measurements with UV illumination on the sample. For both cases we see significant improvement in on/off ratio. For the short term measurements, we see significantly more distinctive current levels associated with each pulse, i.e. the slope of the $I_\text{ds}$ vs. time curve is significantly larger. Because of this, each of the 100 pulses can be uniquely distinguished by the current level in the UV case. For the long term case, we see a large improvement in the current level after the spike when UV is applied. For the non-UV case the postsynaptic current on/off ratio is $\sim$1.01. While when the memtransistor is under UV illumination, this on/off ratio becomes $\sim$1.05. However, in terms of long term stability, we see that the potentiation is only temporary while under continuous UV illumination, due to the discharging caused by the UV light. Ideally the UV should only be applied at the same time as the gate pulse. This synchronization between UV light and gate voltage will be studied in more details in a follow up work.

Finally, we can estimate the energy used for our potentiation/depression (write/erase) of the artificial neuron. Using the 3\,V gate spikes as above, estimating the active device area as $\sim5\times5$\,$\upmu$m$^2$ and using the standard equation for energy charged in capacitor:
\begin{align}
    U_c = \frac{\epsilon_r \epsilon_0 A V_g^2}{2 d},
\end{align}
where $\epsilon_r$ is the relative permittivity of the gate oxide between the graphene and the nitride and $\epsilon_0$ is the vacuum permittivity. $A$ is the device area and $d$ is the distance between the graphene and the nitride. Assuming a 6\,nm silicon dioxide gate we get a write/erase energy of $\sim215$\,fJ. Here we have used the external voltage of 3\,V, however, based on the amount of charges stored in the nitride this voltage will be partially screened, so this number is only an upper estimate of energy used in the spiking of the neuron.

The exact power consumption for the read operation will vary based on the exact conductivity of the graphene channel, but we can estimate it as roughly $\sim2-4$\,$\upmu$W if using the 0.1\,V drain-source voltage as above.

\section*{Discussion}

We see generally  that applying a negative gate bias results in a much stronger and faster change in the graphene's conductivity. Breakdown of the tunnel oxide is also much more likely to occur for larger negative biases. It is for this reason that for the memory experiment we apply asymmetric (in terms of magnitude) pulses of -5\,V and 8\,V, as even applying bias in excess of -15\,V was observed to lead to a dielectric breakdown, while breakdown for the positive bias is usually observed only above 24\,V.

We have here only studied the electronic read-out of the memory state, but as the doping level of graphene has critical influence on its optical properties\cite{Xiao:2016}, a similar device with patterned graphene would allow for optical readout by the shift of a graphene plasmon resonance\cite{Xiao:2016}, or alternatively the graphene layer could be integrated with a waveguide to modify the transmission coefficient of the waveguide\cite{Ansell:2015}.

For the neuropmorhic measurements, we are required to use relatively large and long gate pulses/spikes in order to see clear shifts in the postsynaptic current. However, by how generally every aspect of the measurement is improved during UV illumination, we are optimistic that further optimization of the oxide thicknesses in the ONO stack could lead to much improved device performance. This is because the UV light effectively serves the role to artificially lower the tunneling barrier in the ONO stack, which could also be achieved by more careful device design. Secondly, the length of the applied pulses are largely a limitation of our custom built measurement system, where we cannot perform measurements with much greater than 500\,ms time resolution. However, tunneling processes can normally occur within a few picoseconds, and therefore it is expected that a device such as this could operate at MHz or GHz switching rates.

For power consumption, minimizing the surface area of the graphene/ONO stack could serve to further reduce the energy needed for spiking the articifical neuron to less than the $\sim215$\,fJ we reported here, as the main contribution for the 

\section*{Conclusion}

We have demonstrated here how graphene can be integrated with conventional flash memory technology relying on charge trapping in an ONO stack. We have also shown how this integration results in programmable doping of the graphene. i.e., by applying a gate pulse the graphene's conductivity can be set to either a higher or a lower conductive state. Furthermore, we have demonstrated how the doping change has a great deal of control and have shown $\sim$100 uniquely different conductive states of the graphene channel in a short term neuroplastic measurement. In a long term measurement, we have also shown how these conductive changes remain for several minutes, and they are expected to be near-permanent - corresponding to the nitride layer's capability to retain stored charge (conventional flash memory has nominal charge loss over several years\cite{grajower:2018}).

By the application of UV we have shown how the efficiency of the charge trapping process can be increased, resulting in larger total doping changes of the graphene with the same gate potential and gate duration in the neuromorphic measurements.

Finally, we have also shown that the nitride can be discharged via long-term UV illumination. This can offer a kind of 'factory reset' for a neural network, which allows for a complete reset of all the neurons to their initial state. This could have applications in small neural circuits that need to be quickly be repurposed for several different tasks, or could also be used in studies of how a neural network learns, by resetting the network and observing the learning process several times on the same hardware.

While the demonstrated device has immediate applications as an electronic memory device, it could also be exploited in combination with many other emerging graphene based technologies, such as tunable graphene plasmonics\cite{Xiao:2016}, graphene based light modulators\cite{Ansell:2015}, and molecular sensors\cite{Rodrigo:2015,Xiao:2016}. Additionally, the device architecture presented here, due to its ease of scalability, should also be highly suitable for large neuromorphic computing structures featuring a network of multiple neurons\cite{Sangwan:2018,Sangwan:2020}.


\section*{Methods}
\subsection{Sample fabrication:} The base substrate for our fabrication is 2" p-type Si wafers (0.02 - 0.04\,$\Upomega$cm resistivity). The first step is to define the individual device areas, where the ONO stack will be confined within. This is done by a standard lithography process using n-LOF 2020 resist and a laser writer system.

Next, the ONO stack is fabricated. The bottom silicon oxide layer is grown by thermal dry oxidation in a 920\,$^\circ$C furnace. Then silicon nitride is grown using PECVD at 300\,$^\circ$C for 1\,min. The sample is then cleaned in piranha (H$_2$SO$_4$:H$_2$O$_2$ ; 3:1), and the nitride layer is wet oxidized at 1000\,$^\circ$C to form the top oxide layer. The final stack has 6\,nm bottom oxide, 4\,nm middle nitride, and 6\,nm top oxide.

An aluminum backgate electrode is then sputtered on the back side of the wafer, and alloyed to the silicon to form an Ohmic contact by heating the device to 460\,$^\circ$C in N$_2$ atmosphere. Gold contacts are defined by a lithography step on the top surface to form the source and drain electrodes.

Finally, the graphene layer which has been prepared by exfoliation from natural bulk graphite crystals (NGS Naturgraphit), is dry transferred by picking it up with an exfoliated layer of bulk hexagonal boron nitride (hBN, HQ graphene) using a viscoelastic polymer transfer method\cite{pizzocchero:2016}. The graphene/hBN stack is then placed so that it straddles the source and drain electrodes across the ONO channel.

\subsection{Measurement:} The devices were measured electrically using a micro-mechanical probe system. Tungsten probes were mechanically pressed against the gold electrodes on the top of the device using XYZ 500MIM 3D stages (Quater Research \& Development), and the backgate electrode was shorted to the sample stage and electrically connected. The sample was attached to the stage using vacuum to ensure a good connection. The gate bias was applied using a Keysight B2901A source meter, while the $I_\text{ds}$ current was monitored using a Keithley 2400 source meter, applying a constant bias of 0.1\,V between the source and drain. All the measurements were performed in air at room temperature. The two source meters were controlled simultaneously by a custom MATLAB script using the Instrument Control Toolbox via a GBIP connection.

UV light was achieved by illuminating the entire sample with a 3\,W LightFe UV301D Flashlight, with peak emission at 365\,nm.

\vspace{12pt}

\subsection{Acknowledgments:} 
We acknowledge funding from the Israeli Ministry of Science and Technology and The Air Force Office of Scientific Research.  C.F. is supported by the Carlsberg Foundation as an Internationalisation Fellow. 
The authors would like to thank Dr. Devidas Taget Raghavendran, Ayelet Zalic and Prof. Hadar Steinberg from the Racah institute of Physics at HUJI. We also acknowledge  Atzmon Vakahi and Dr. Sergei Remennik from Harvey M. Kruger Family Center of Nanoscience and Technology for assistance with the STEM imaging and FIB sample preparation. 

\subsection{Conflicts of interests:} 
The authors declare no competing financial interests.

\subsection{Correspondence:} General correspondence should be addressed to: stiaram.indukuri@mail.huji.ac.il or ulevy@mail.huji.ac.il

\bibliography{main}

\end{document}